# Optimal conditions for the generation of moderate-order harmonics of a short-wave field by helium atoms

V.A. Antonov[1,*], I.R. Khairulin[1,2], M.Yu. Emelin[1], M.M. Popova[1,3], E.V. Gryzlova[1,3], and M.Yu. Ryabikin[1,2]

[1]*Gaponov-Grekhov Institute of Applied Physics of the Russian Academy of Sciences, 46 Ulyanov Street, Nizhny Novgorod 603950, Russia*
[2]*Lobachevsky State University of Nizhny Novgorod, 23 Gagarin Avenue, Nizhny Novgorod 603950, Russia*
[3]*Skobeltsyn Institute of Nuclear Physics, Lomonosov Moscow State University, 119991 Moscow, Russia*

[*]Corresponding author: V.A. Antonov, antonov@appl.sci-nnov.ru

It is shown that under optimal conditions, the generation of the 3rd, 5th, 7th, and 9th harmonics of the short-wave, ultraviolet or vacuum ultraviolet, laser field by helium atoms is mainly due to transitions between bound states, and the maximum energy of the harmonics is achieved under conditions of their resonant multiphoton excitation. In this case, the optima for the generation of the 3rd, 5th, 7th, and 9th harmonics of the field correspond to three-, four-, five-, and six-photon resonances, and the optimal value of the peak field intensity, depending on the harmonic order, varies from $2.5\times10^{14}$ W/cm$^2$ to $1.2\times10^{15}$ W/cm$^2$. The total probability of excitation and ionization of an atom at the end of a laser pulse under corresponding conditions exceeds 1/2. With such intensity and not-too-high frequency of the laser field, the Stark effect turns out to be very significant, which allows it to be tuned in resonance with an arbitrary excited state of the atom by changing the intensity of the field without changing its frequency. It is shown that for the laser field parameters maximizing the energy of the $N$th harmonic, $3 \le N \le 9$, this harmonic dominates in the dipole acceleration spectrum. At the same time, the amplitudes of the remaining harmonics increase as the harmonic order approaches $N$. In particular, under the conditions maximizing the yield of the 9th harmonic, the harmonic amplitudes increase when moving from the 3rd harmonic to the 5th and then to the 7th and 9th harmonics, and the harmonics form an attosecond pulse train. At the same time, under the conditions maximizing the yield of the 3rd harmonic, its amplitude in the dipole acceleration spectrum exceeds not only the amplitudes of the other harmonics, but also the amplitude of the atomic response at the frequency of the driving field.

## I. INTRODUCTION

The development of lasers in the 1960s led to technological advances that further required the reduction of laser wavelengths to (a) increase the spatial and temporal resolution available with these sources, (b) drive higher energy processes (such as electron excitation from inner shells), (c) reduce the effects of plasma on radiation propagation, and (d) increase its peak power.

To date, various methods have been proposed for generating coherent short-wave radiation, of which the generation of laser field harmonics provides the highest degree of coherence. In addition, such sources are compact and widely available, which increases their practical value.

Historically, the study and practical application of laser field harmonics began in the 60s and 70s of the XXth century with harmonics of lower orders – the second and third, generated in a relatively weak field and described by perturbation theory (in this case, the amplitude of the harmonic of each subsequent order is much lower than the amplitude of the harmonic of the previous order) [1-3]. An important advantage of harmonics of lower orders is the high efficiency of their generation, which for the second harmonic can exceed 50% [4-6].

The next class of laser field harmonics to be studied was high-order harmonics generated by a three-step (Corkum's) mechanism that includes ionization of an atom, acceleration of a released electron in a laser field, and, upon returning to the atomic core, its subsequent recombination with the parent ion. Harmonics of this class were discovered at the turn of the 80s and 90s of the XXth century [7-9], and their study is still ongoing [10, 11]. As in the case of low-order harmonics, the generation of high harmonics is correctly described by perturbation theory, but in its inverse limit, namely, for a strong laser field and weak interaction of an electron with the atomic core. In this case, a plateau is formed in the spectrum of generated harmonics, which is a region of photon energies significantly exceeding the ionization potential of the medium, within which the harmonic amplitude weakly depends on its order. In addition, the phases of the harmonics in the plateau region can be ordered. Together, these factors make it possible to generate harmonic radiation with a total bandwidth of up to 12 octaves and photon energy of up to 1.6 keV [12], and to form pulses from high harmonics with a duration of less than 50 attoseconds [13], see also [10, 11]. At the same time, the main disadvantage of high-order harmonics, which limits their practical applications, is the relatively low power, due to the low generation efficiency, which in typical experimental conditions varies from $10^{-4}$ to $10^{-9}$ depending on the harmonic order.

In this regard, in recent years considerable attention has been paid to the generation of below-threshold and near-threshold harmonics of moderate orders with photon energies lower than or comparable to the ionization potentials of neutral media [14-27]. As recent experiments and numerical simulations show, the typical efficiency of generating such harmonics is much higher than that of high-order harmonics, see, for example, [16]. At the same time, the closeness of photon energies to excitation energies and ionization potentials of neutral media makes such radiation an extremely effective tool for controlling their quantum state. However, the description of the process of generating moderate-order harmonics is a complex problem that cannot be solved by perturbation theory, since under typical conditions the field acting on the outer electron from the atomic core is of the same order of magnitude as the laser field. As a result, the properties of such harmonics have not been fully understood. Thus, there is no complete picture of the optimal conditions for generating below- and near-threshold harmonics of different orders, and the spectral and temporal properties of sets of such harmonics have been poorly studied. It should also be noted that most previous studies on the generation of moderate-order harmonics were related to the case of a Ti:Sa or longer wavelength driving field with a wavelength of 0.6 to 3.6 μm mostly in the infrared (IR) range. For noble gas atoms, such a field is long-wave (the photon energy is an order of magnitude smaller than the excitation energies and ionization potentials of the atoms), which reduces the role of multiphoton resonances, and, consequently, increases the role of the rescattering of a released electron (the Corkum's mechanism) in the moderate-order harmonic generation.

In contrast to the above-mentioned works, this article discusses the generation of moderate-order harmonics of a short-wave, predominantly ultraviolet (UV) and vacuum ultraviolet (VUV), laser field with a wavelength from 91.1 nm to 570 nm. This spectral range covers unperturbed frequencies from two-photon to ten-photon transitions from the ground to bound excited states of the helium atom. For such a short-wave laser field, the role of multiphoton resonances greatly increases, which, as will be shown below, makes them the dominant mechanism for the generation of moderate-order harmonics. We study the harmonic generation by helium atoms in a given external field. This case corresponds to an extremely thin medium for which, regardless of its degree of ionization and concentration of free electrons, any propagation effects can be neglected. The main attention is paid to the optimal conditions for generating the third through ninth harmonics of the field.

The paper is organized as follows. Introduction (Sec. I) is followed by a description of the theoretical models used (Sec. II). Section III presents a general analysis of the properties of the solution within the two-dimensional (2D) one-electron model of the helium atom. Section IV compares the results of the 2D one-electron model with the calculations performed by expanding the wave function of the helium atom over the basis of two-electron stationary states. Optimum conditions for generating the 3rd, 5th, 7th, and 9th harmonics are discussed. Section V analyzes the influence of resonance order on the efficiency of generating moderate-order harmonics. Section VI analyzes the spectra of moderate-order harmonics, the time dependences of induced dipole acceleration, and the population dynamics of stationary states under conditions of harmonic energy maximization. Section VII concludes the paper.

## II. THEORETICAL MODELS

Let us consider the generation of harmonics of a linearly polarized short-wave laser field by helium atoms. Let's define the electric part of the laser field in the form

$$\vec{E}(t) = \vec{z}_0 E_0 f(t)\sin(\Omega t), \tag{1}$$

where $\vec{z}_0$ is a unit vector along the $z$ axis, the direction of polarization of the laser field, $E_0$ and $\Omega$ are the amplitude and carrier frequency of the field, respectively, and $f(t)$ is its envelope. Further we will assume that $f(t)$ has the form of a square sine with a full duration (from zero to zero) $\Delta t_0$ :

$$f(t) = \left[\theta(t-\Delta t_0) - \theta(t)\right]\sin^2\left(\pi t/\Delta t_0\right), \tag{2}$$

where $\theta(t)$ is the Heaviside unit step function. In what follows, we will consider the frequency range $\Omega$ from 0.08 to 0.5 atomic units (a.u.), corresponding to the laser wavelengths in vacuum from 91.1 to 570 nm and covering the unperturbed frequencies of multiphoton transitions of orders from the 2nd to the 10th from the ground to the excited bound states of the helium atom. We will assume that the total duration of the field pulse is equal to 82 periods of its oscillations, $\Delta t_0 = 82T$ , where $T = 2\pi/\Omega$. In this case, the full width at half maximum (FWHM) of the laser pulse, $\Delta t_{1/2} = 0.364\Delta t_0$, is approximately equal to 30 periods of field oscillations and, with a change in its frequency, changes from 9.1 fs for $\Omega$ = 0.5 a.u. to 57 fs for $\Omega$ = 0.08 a.u. In turn, the intensity of the laser field, $I_0 = cE_0^2/(8\pi)$ (where $c$ is the speed of light in vacuum), varies in the range from $10^{13}$ W/cm$^2$, which corresponds to the perturbative excitation regime of the atom (the weak field limit), to $10^{16}$ W/cm$^2$, which corresponds to almost complete ionization of the atom by the laser pulse (the strong field limit).

We will assume that the harmonic emission occurs in an ultimately thin layer of helium atoms, so that the effects of propagation can be neglected. In this case, the electric field of the generated radiation is proportional to the induced dipole acceleration of the atom, i.e., the second derivative with respect to time of the induced dipole moment:

$$\ddot{d}(t) = \frac{d^2}{dt^2}\left\langle \Psi(t)\right| \hat{d}_z \left|\Psi(t)\right\rangle, \tag{3}$$

where $\hat{d}_z$ is the projection of the dipole moment operator of the atom onto the direction of polarization of the laser field (1) and $\left|\Psi(t)\right\rangle \equiv \left|\Psi(\vec{r}_1,\vec{r}_2,t)\right\rangle$ is the two-electron wave function of the helium atom satisfying the time-dependent Schrödinger equation (TDSE)

$$i\frac{\partial}{\partial t}\left|\Psi(\vec{r}_1,\vec{r}_2,t)\right\rangle = \left[\hat{H}_0 + \hat{d}_z \cdot E(t)\right]\left|\Psi(\vec{r}_1,\vec{r}_2,t)\right\rangle, \tag{4}$$

in which $\vec{r}_i$, $i$ = 1,2, are the coordinates of the electrons in the atom, $\hat{H}_0$ is the Hamiltonian of the atom in the absence of an external field, and before interaction with the laser field the atom is in the ground state $|1s^2\rangle$.

In the following, we are interested in the generation of below-threshold and near-threshold harmonics with photon energies significantly lower than the energies of two-electron excitations of the helium atom, i.e. the autoionization states (the excitation energies of which exceed 58.3 eV) and the two-electron continuum (with the photoionization threshold of 79 eV). This allows us to ignore two-electron excitations (affecting only harmonics of considerably higher photon energies, see, for example, [28, 29]), when solving the TDSE (4) and subsequently calculating the atomic dipole acceleration (3).

To solve Equation (4), we used two approaches. The first one consists of a numerical solution of the TDSE from first principles in a 2D one-electron model of the He atom with an effective potential that reproduces the characteristics of the lowest energy levels of the real atom (for details, see Appendix A in [30]):

$$U(x,z) = -\frac{1+(1+8.125r)e^{-8.125r}}{\sqrt{r^2+0.01}} + \frac{0.6r^6}{r^8+10^{-4}}, \tag{5}$$

where $r=\sqrt{x^2+z^2}$ . In this case, the dipole acceleration for an atom is calculated using the Ehrenfest's theorem:

$$\ddot{d}_{\text{2D}}(t) = E(t) + \int \Psi^* \frac{\partial U}{\partial z} \Psi dxdz, \tag{6}$$

and the probability of atomic ionization is calculated as

$$W_{ion}^{(\text{2D})}(t) = 1 - \int |\Psi|^2 \, dxdz, \tag{7}$$

where the integration is performed within a square region with a side of 100 a.u. centered at the atomic nucleus, and the index "2D" for the dipole acceleration and ionization probability indicates the two-dimensional one-electron model within which they are calculated. Note that the wave functions of the atomic bound states with the principal quantum numbers $n$ = 1,2,3,4 fit into the spatial grid used to solve the Schrödinger equation without significant distortions. Together with the careful choice of the effective 2D one-electron potential in the form (5) this made it possible to reasonably reproduce the characteristics of the bound states of a real helium atom with a principal quantum number up to $n$=4, which makes it suitable for describing the multiphoton transitions between them. The main advantage of this approach is the possibility of jointly taking into account the laser-induced one-electron transitions between the He atom states of all possible types, namely, transitions between bound states, from bound to free states and back, as well as transitions between continuum states. However, the parameters of states in the 2D one-electron model generally differ from the parameters of a real 3D two-electron helium atom. In addition, the limited size of the spatial grid used in the numerical integration of the TDSE prevents the correct description of highly excited bound states, which limits the accuracy of the description of bound-bound transitions.

The second approach we used to solve the TDSE is to expand the two-electron wave function of the helium atom in a basis of stationary states, including the ground state of the atom and a limited, but sufficiently large, number of its one-electron excitations. Namely, the expansion takes into account the bound states of the atom and the states of the single-electron continuum into which electric dipole transitions from the considered bound states are allowed:

$$\left|\Psi(\vec{r}_1,\vec{r}_2,t)\right\rangle = \sum_{k=1}^{K_{\max}} a_k(t)\left|\psi_k(\vec{r}_1,\vec{r}_2)\right\rangle + \sum_{l=0}^{L_{\max}} \int_0^{\infty} d\varepsilon \cdot b_l(\varepsilon,t)\left|\varepsilon,l\right\rangle. \quad (8)$$

Here $a_k(t)$ is the excitation amplitude of the bound stationary state $\left|\psi_k(\vec{r}_1,\vec{r}_2)\right\rangle$ and $b_l(\varepsilon,t)$ is the excitation amplitude of the continuum state $\left|\varepsilon,l\right\rangle$ with energy $\varepsilon>0$ and orbital momentum *l*, $0\le l\le L_{\max}$. In the calculations presented here, $K_{\max}$ = 36 bound states of the He atom with principal quantum numbers $n$ = 1,…,8, as well as continuum states with energies $0 < \varepsilon \le 1.5$ a.u. and orbital momentum $l$ from 0 to $L_{\max}$ = 8 were taken into account in the expansion of the wave function. For all states under consideration, the projection of the orbital momentum of the atom onto the polarization axis of the laser field (1) is $m$ = 0, since electric dipole transitions from the atomic ground state $\left|1s^2\right\rangle$ to states with $m\neq 0$ in a linearly polarized field are forbidden by the selection rules. Substituting expansion (8) into (4) and using the method of adiabatic elimination of continuum states in the rotating wave approximation [31], we obtain the following set of equations for the excitation amplitudes $a_k(t)$:

$$\frac{da_k}{dt} = -\left[i\mathrm{E}_k + \gamma_k(t)\right]a_k - iE(t)\sum_{s=1}^{K_{\max}} d_{sk}^{(z)} a_s,\ k=1,\ldots,K_{\max}, \quad (9)$$

where $\mathrm{E}_k<0$ is the energy of the bound state $\left|\psi_k(\vec{r}_1,\vec{r}_2)\right\rangle$, $d_{sk}^{(z)} = \left\langle\psi_k\right|\hat{d}_z\left|\psi_s\right\rangle$ is the *z*-projection of the matrix element of the dipole moment operator of the transition from the bound state $\left|\psi_s(\vec{r}_1,\vec{r}_2)\right\rangle$ to the bound state $\left|\psi_k(\vec{r}_1,\vec{r}_2)\right\rangle$, and $\gamma_k(t)$ is the rate of depletion of the $\left|\psi_k(\vec{r}_1,\vec{r}_2)\right\rangle$ state due to single-photon ionization by the laser field (1):

$$\gamma_k(t) = \frac{\pi}{4}E_0^2 f^2(t)\sum_{l=0}^{L_{\max}}\left|\left\langle \mathrm{E}_k+\Omega,l\right|\hat{d}_z\left|\psi_k\right\rangle\right|^2 \theta\left(\mathrm{E}_k+\Omega\right), \quad (10)$$

where $\left\langle \mathrm{E}_k+\Omega,l\right|\hat{d}_z\left|\psi_k\right\rangle$ is the *z* projection of the matrix element of the dipole moment operator of the transition from the bound state $\left|\psi_k\right\rangle$ to the continuum state $\left|\mathrm{E}_k+\Omega,l\right\rangle$. The set of equations (9) was solved with the initial conditions

$$a_1\left(t\to-\infty\right)=1,\ a_k\left(t\to-\infty\right)=0 \ \text{if}\ \ k\neq 1, \quad (11)$$

where $a_1(t)$ is the amplitude of the atomic ground state $\left|1s^2\right\rangle$. The energies of the bound states and the dipole moments of the transitions included in (9) and (10) were calculated by numerically solving the 3D two-electron stationary Schrödinger equation using the multiconfiguration Hartree-Fock method [32], which made it possible to reproduce with high accuracy the characteristics of a real 3D two-electron helium atom. The main limitation of the approach (8)-(10) is the consideration of the continuum states as a “thermostat” into which an active electron can pass as a result of ionization, but does not return back. Thus, this approach does not take into account transitions from continuum to bound states (recombination), as well as transitions between continuum states. At the same time, the correct consideration of a larger number of bound states of an atom compared to the 2D single-electron model makes it more accurate in describing bound-bound transitions. Besides, this approach allows one to analyze the time dynamics of populations of stationary states of an atom in a field.

When using Eqs. (8)-(10), the induced dipole acceleration of the atom is expressed through the excitation amplitudes $a_k(t)$ as follows:

$$\ddot{d}_{\exp}(t) = \frac{d^2}{dt^2} \sum_{k,s=1}^{K_{\max}} d_{sk}^{(z)} a_k^*(t) a_s(t), \tag{12}$$

where the index "exp" denotes the method of calculating the desired quantities through expansion of the atomic wave function in stationary states. It should be noted that, unlike $\ddot{d}_{2\mathrm{D}}(t)$, in this case, when calculating the dipole acceleration, the states of the continuum are not taken into account and, accordingly, there are no contributions from free-free and free-bound transitions (Brunel's [33] and Corkum's [8, 9] harmonic generation mechanisms).

Note that the excitation amplitudes of stationary states $a_k(t)$ and the dipole acceleration of the atom can be expanded in Fourier series in terms of the harmonics of the laser field. In this case, in a not-too-strong laser field (under the condition of slow excitation and ionization of the atom on the scale of the laser cycle) with a slowly varying envelope $f(t)$ (2), the expansion coefficients are slow functions of time:

$$a_k(t) = \sum_{p=-\infty}^{\infty} a_{k,p}(t) e^{-ip\Omega t},\ k = 1,\ldots,K_{\max},\quad \left|\frac{d}{dt} a_{k,p}(t)\right| \ll \Omega \left|a_{k,p}(t)\right|, \tag{13a}$$

$$\ddot{d}_{\exp}(t) = \sum_{N=-\infty}^{\infty} \ddot{d}_{\exp}^{(N)}(t) e^{-iN\Omega t},\quad \left|\frac{d}{dt} \ddot{d}_{\exp}^{(N)}(t)\right| \ll \Omega \left|\ddot{d}_{\exp}^{(N)}(t)\right|. \tag{13b}$$

Combining (12) and (13), we obtain

$$\ddot{d}_{\exp}^{(N)}(t) = -\left(N\Omega\right)^2 \sum_{p=-\infty}^{\infty} \sum_{k,s=1}^{K_{\max}} d_{sk} a_{k,p}^* a_{s,p+N}. \tag{14}$$

In turn, the probability of ionization of an atom in this case is calculated as

$$W_{ion}^{(\exp)}(t) = 1 - \sum_{k=1}^{K_{\max}} \left|a_k(t)\right|^2. \tag{15}$$

Finally, to estimate the efficiency of the *N*-th order harmonic generation, we use the harmonic energy value defined as the integral of the spectral power density of the dipole acceleration in the vicinity of the harmonic frequency:

$$\mathrm{E}_N^{(2\mathrm{D})} = \frac{1}{2\pi} \int_{(N-1)\Omega}^{(N+1)\Omega} \left|s_{2\mathrm{D}}(\omega)\right|^2 d\omega, \tag{16a}$$

$$\mathrm{E}_N^{(\exp)} = \frac{1}{2\pi} \int_{(N-1)\Omega}^{(N+1)\Omega} \left|s_{\exp}(\omega)\right|^2 d\omega, \tag{16b}$$

where $s_{2\mathrm{D}}(\omega) = \int_{-\infty}^{\infty} \ddot{d}_{2\mathrm{D}}(t) f(t) e^{i\omega t} dt$ and $s_{\exp}(\omega) = \int_{-\infty}^{\infty} \ddot{d}_{\exp}(t) f(t) e^{i\omega t} dt$ are the spectra of the dipole acceleration of the atom calculated using a temporal mask that coincides with the envelope of the laser field. The spectrum of harmonics calculated in this way allows one to minimize the influence of radiation at the natural atomic frequencies.

## III. ANALYSIS OF THE RESULTS FOR A TWO-DIMENSIONAL SINGLE-ELECTRON MODEL OF THE HELIUM ATOM

Let us consider Fig. 1, which illustrates the dependences of the energy of the 3rd, 5th, 7th, and 9th harmonics in the dipole acceleration spectrum of the helium atom on the laser frequency for different values of its peak intensity (neighboring intensity values in Fig. 1 differ by a factor of two). This figure is drawn based on the solution of the 2D one-electron TDSE from first principles. It is evident from this figure that, firstly, the frequency dependence of the energy of each harmonic is a set of resonant peaks at the frequencies of multiphoton transitions from the ground to the bound excited atomic states. Secondly, with increasing intensity of the laser field, the resonances shift to the high-frequency region due to the Stark effect. Thirdly, the energies of all the considered harmonics, with the exception of the 3rd, reach a global maximum at a peak field intensity of $10^{15}$ W/cm$^2$, whereas the energy of the third harmonic reaches a global maximum at a peak intensity of $5\times10^{14}$ W/cm$^2$. Fourthly, the maximum achievable values of the energy of harmonics of different orders are comparable. In particular, the peak energies of the 3rd and 9th harmonics, achieved under conditions optimal for their generation, differ by less than an order of magnitude.

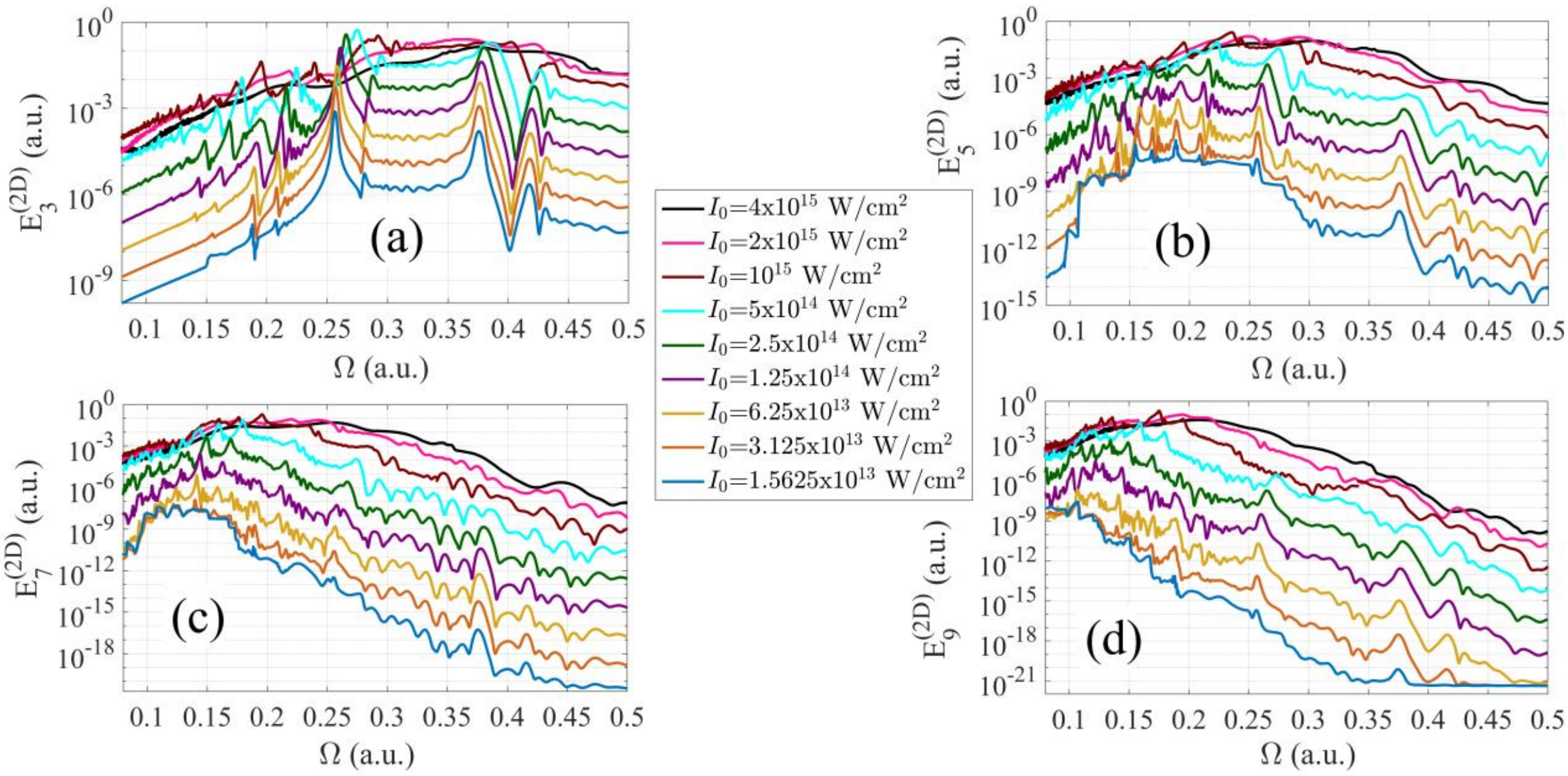

FIG. 1. Dependences of the energies $E_N^{(2D)}$ of harmonics of order (a) $N = 3$, (b) $N = 5$, (c) $N = 7$, and (d) $N = 9$ on the laser field frequency Ω at different peak intensity values $I_0$ (see central inset), obtained on the basis of a numerical solution of the TDSE in a 2D one-electron model of the helium atom.

In order to follow the change in the frequencies of resonant multiphoton transitions with a change in the intensity of the laser field, we turn to Fig. 2, which shows the dependence of the ionization probability of a helium atom, $W_{ion}^{(2D)}$, after the end of the laser pulse on its carrier frequency Ω and peak intensity $I_0$. In this figure, in addition to the results of solving the 2D one-electron TDSE from first principles, estimates of the laser field frequency are given, which correspond to the upper limit of the frequencies of $M$-photon transitions from the atomic ground state $|1s^2\rangle$ to excited bound states (black curves in Fig. 2), i.e.

$$\Omega_{p\,\mathrm{shifted}}^{(M)} = \frac{I_p}{M} + \frac{E_0^2}{4M\left(\Omega_{p\,\mathrm{shifted}}^{(M)}\right)^2}, \qquad (17)$$

where $I_p$ is the ionization potential of the He atom (for the 2D one-electron model $I_p = 0.9034$ a.u., while for the model based on the expansion in the basis of stationary states $I_p = 0.9036$ a.u., see Appendix A in [30]), and the second term in Eq. (17) is the ponderomotive energy of a free electron in the laser field. In addition, Fig. 2 shows the estimates of the frequencies of $M$-photon transitions from the atomic ground state $|1s^2\rangle$ to excited states $|1s2s\rangle$ ($M$ is even, cyan curves in Fig. 2) and $|1s2p\rangle$ ($M$ is odd, light green curves in Fig. 2), respectively:

$$\Omega_{\mathrm{1s2s}}^{(M)} = \frac{\omega_{\mathrm{1s2s}}}{M} + \frac{E_0^2}{4M\left(\Omega_{\mathrm{1s2s}}^{(M)}\right)^2}, \tag{18a}$$

$$\Omega_{\mathrm{1s2p}}^{(M)} = \frac{\omega_{\mathrm{1s2p}}}{M} + \frac{E_0^2}{4M\left(\Omega_{\mathrm{1s2p}}^{(M)}\right)^2}, \tag{18b}$$

where $\omega_{\mathrm{1s2s}}$ and $\omega_{\mathrm{1s2p}}$ are the unperturbed frequencies of the transitions $|1\mathrm{s}^2\rangle \leftrightarrow |1\mathrm{s2s}\rangle$ and $|1\mathrm{s}^2\rangle \leftrightarrow |1\mathrm{s2p}\rangle$, respectively. These estimates assume that in the laser field the energy of the $|1\mathrm{s}^2\rangle$ state does not change, whereas the ponderomotive energy of a free electron is added to the energies of the excited states due to the Stark effect.

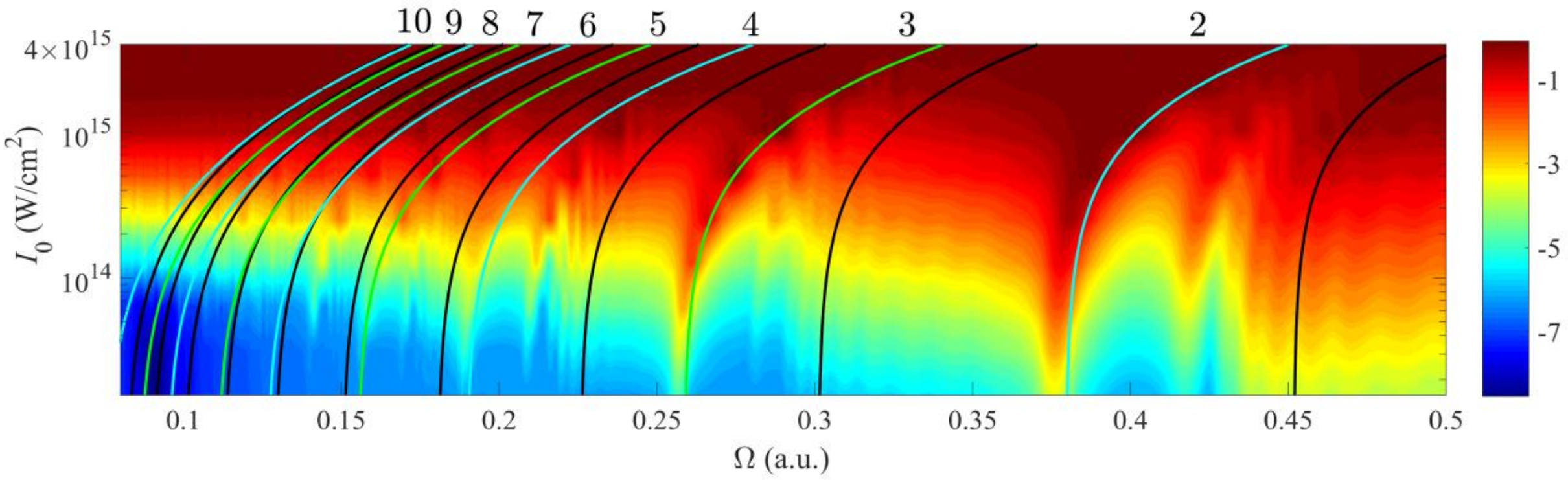

FIG. 2. Dependence of the decimal logarithm of the ionization probability of a helium atom, $\log_{10}(W_{ion}^{(2D)})$, at the end of the laser pulse (1), (2) on its frequency Ω and peak intensity $I_0$, calculated in the 2D one-electron model. Black curves denote estimates (17) for the boundaries of multiphoton transitions of orders $M = 2,\ldots,11$; cyan and light green curves show estimates (18) for the frequencies of transitions from the ground state, $|1\mathrm{s}^2\rangle$, to the lowest atomic excited states, $|1\mathrm{s2s}\rangle$ and $|1\mathrm{s2p}\rangle$, of even and odd orders, respectively, starting with 2 (the far right cyan curve) and ending with 10 (the far left cyan curve).

As follows from Fig. 2, estimates (18) qualitatively correctly reproduce the positions of multiphoton resonances in the solution of the Schrödinger equation from first principles. For the frequencies of three- and five-photon transitions $|1\mathrm{s}^2\rangle \leftrightarrow |1\mathrm{s2p}\rangle$, this estimate turns out to be more accurate than for two- and four-photon transitions $|1\mathrm{s}^2\rangle \leftrightarrow |1\mathrm{s2s}\rangle$. At the same time, for resonances of higher orders, $M \geq 6$, there is an overlap of the frequencies of $M$-photon transitions to lower bound states $|1\mathrm{s2s}\rangle$ and $|1\mathrm{s2p}\rangle$ (18) with the frequencies of $(M + 1)$-photon transitions to highly excited states located near the ionization potential of the atom (17) (the black and colored curves in Fig. 2 intersect). As a result, the corresponding resonance peaks merge, so that estimates (18) largely lose their meaning, and the separation of resonances of different orders in the 2D one-electron model of the helium atom becomes questionable.

Note that with increasing intensity $I_0$ and decreasing frequency Ω of the laser field, the frequencies of multiphoton transitions rapidly increase as a result of the Stark effect, the magnitude of which in a sufficiently low-frequency field can be comparable with the excitation energies of stationary states. Thus, the radiation of the second harmonic of a Ti:Sa laser with a wavelength of 400 nm and a photon energy of 3.1 eV (0.114 a.u.) in a field with an intensity of $6.25\times10^{13}$ W/cm$^2$ is in the vicinity of the 7-photon resonance with the transition $|1\mathrm{s}^2\rangle \leftrightarrow |1\mathrm{s2p}\rangle$, whereas in a field with an intensity of $10^{15}$ W/cm$^2$ optimal for generating the 5th, 7th, and 9th harmonics, it is in the region of 12-photon resonances (the free-electron ponderomotive energy in this case is about 15 eV). Thus, in a sufficiently low-frequency field with a photon energy much lower than the ionization potential of the atom, the search for optimal conditions for resonantly enhanced generation of moderate-order harmonics is reduced to determining the optimal field intensity. There is no need for tuning of the field frequency, since changing the field intensity allows it to be tuned into resonance with any state of the atom and, moreover, to change the order of resonance.

Note also that, as follows from Fig. 2, when the field peak intensity changes from $1.56\times10^{13}$ W/cm$^2$ to $4\times10^{15}$ W/cm$^2$ for the considered laser pulse duration, we pass from the perturbative excitation of the helium atom, when the ionization probability for any field frequency is close to zero, to the strong-field regime and very essential ionization (exceeding 60% for an arbitrary field frequency). As follows from a comparison of Figs. 1 and 2, for the field parameters that maximize the energy of moderate-order harmonics, the atom ionization probability after the end of the laser pulse is about 50%, and the interaction of the atom with the field turns out to be essentially nonperturbative.

## IV. COMPARISON OF THE MODELS. OPTIMUM CONDITIONS FOR THE GENERATION OF THE 3RD, 5TH, 7TH, AND 9TH HARMONICS

Let us now turn to the results of solving the Schrödinger equation based on the wave-function expansion over the basis of stationary states. Note that this approach is less resource-intensive, which allows us to analyze the optimal conditions for generating all the harmonics under consideration with a higher discretization in frequency $\Omega$ and intensity $I_0$ of the laser field, and also to consider a wider range of intensities (from $10^{13}$ to $10^{16}$ W/cm$^2$ inclusive).

Figure 3 shows the 2D plots of the energies of harmonics of orders 3 to 9 in the spectrum of the atomic dipole acceleration, $\mathrm{E}_N^{(\mathrm{exp})}$, $3\leq N\leq 9$, versus the frequency and peak intensity of the laser field calculated by this method. The points corresponding to the global maxima of the harmonic energy are marked with white asterisks. For comparison, Fig. 4 shows similar dependences for $\mathrm{E}_N^{(2\mathrm{D})}$ calculated in the 2D one-electron model of the helium atom. The laser field parameters that maximize the energies of the 3rd, 5th, 7th, and 9th harmonics in the spectrum of the dipole acceleration of the atom, obtained in each of the models, as well as the corresponding harmonic frequencies, are summarized in Table 1.

First of all, we note that the results of both models of the helium atom are qualitatively similar. In both cases, with an increase in the order of the harmonic $N$, (a) the field peak intensity that maximizes the harmonic yield increases, while (b) the corresponding optimal frequency of the laser field decreases. The frequency of the harmonic itself under conditions optimal for its generation is of the order of or somewhat greater than the excitation energies of the bound states and the atomic ionization potential.

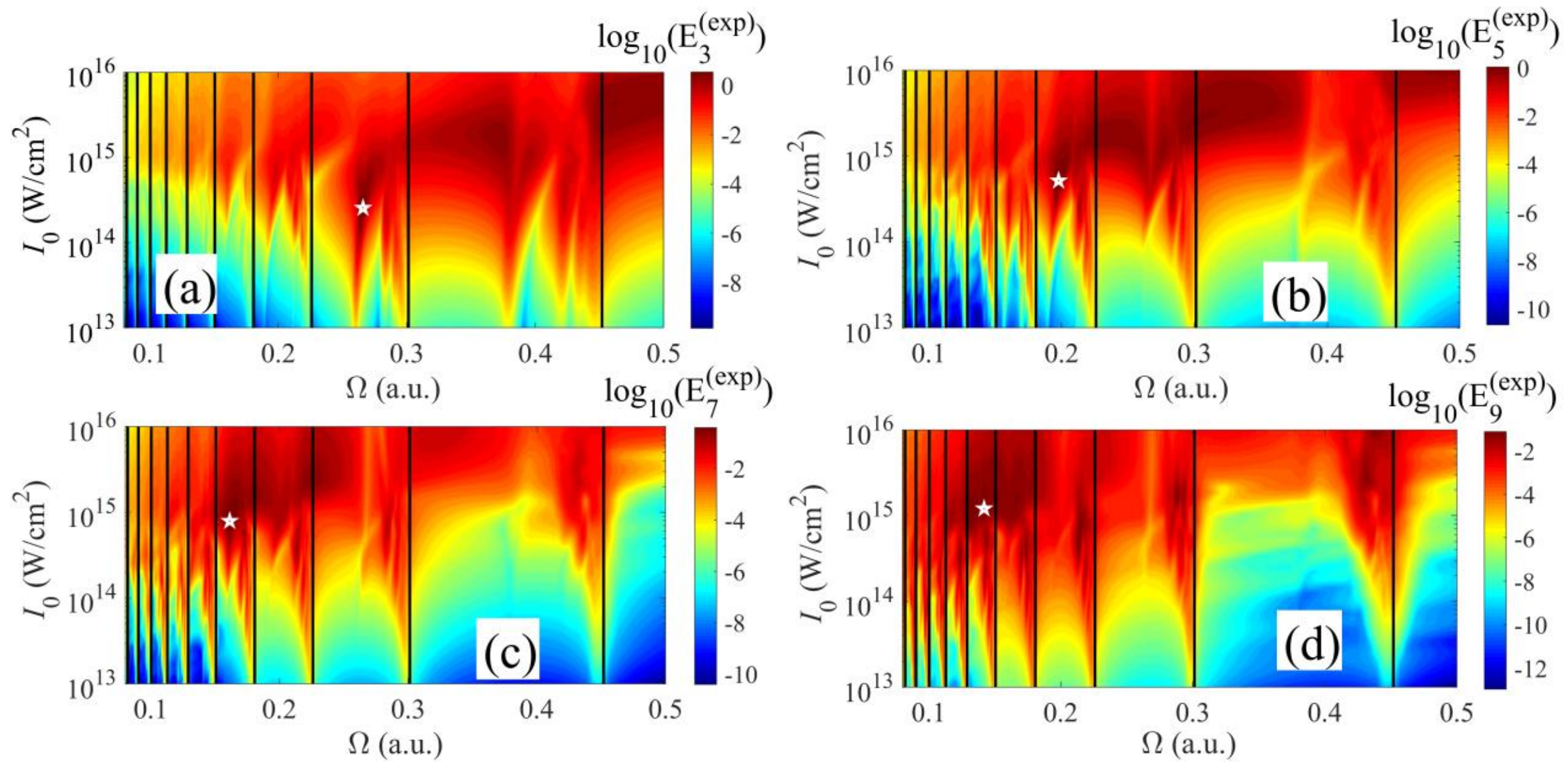

FIG. 3. Dependences of the decimal logarithm of the energies $\log_{10}(E_N^{(\mathrm{exp})})$ of the harmonics of order (a) $N = 3$, (b) $N = 5$, (c) $N = 7$, and (d) $N = 9$ in the dipole acceleration spectrum of the helium atom on the frequency $\Omega$ and peak intensity $I_0$ of the laser field, obtained by expanding the atomic wave function in the basis of stationary states. Black lines indicate the estimates of the boundaries of multiphoton transitions of orders $M = 2,\ldots,11$, determined with the unperturbed value of the atomic ionization potential: $\Omega_p^{(M)} = I_p / M$ . Asterisks mark the global maxima of the energy of the harmonics under consideration (see also Table 1).

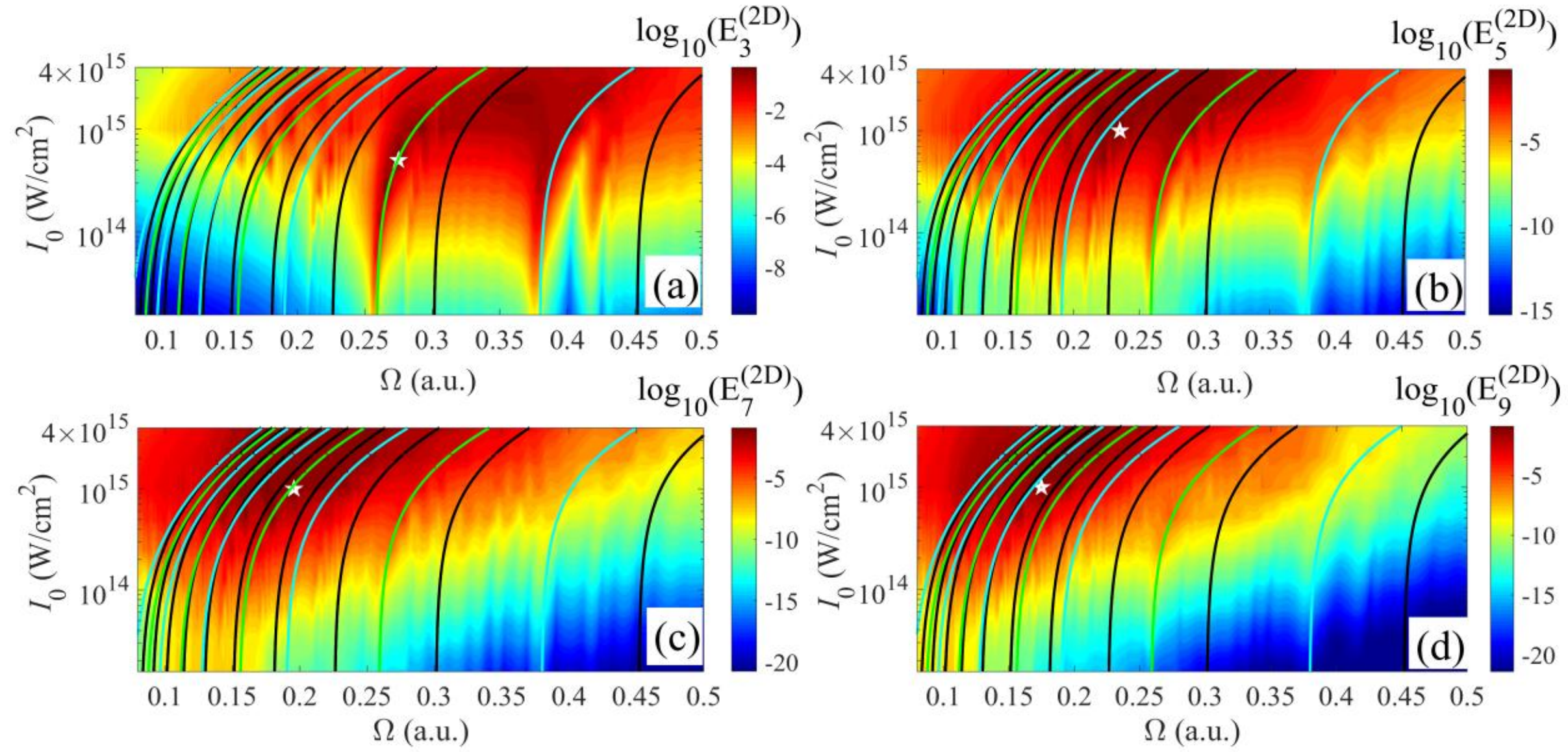

FIG. 4. Frequency and intensity dependences for $\log_{10}(E_N^{(2D)})$, calculated similarly to Fig. 3, but in the 2D one-electron model of the helium atom. Black, cyan, and light green curves have the same meaning as in Fig. 2.

Table 1. The laser field parameters that maximize the energies of the 3rd, 5th, 7th, and 9th harmonics in the spectrum of the dipole acceleration of the atom, and the corresponding harmonic frequencies. Note that the ionization potential of the He atom is $I_p = 0.9034$ a.u. for the 2D one-

electron model and $I_p$ = 0.9036 a.u. for the model based on the expansion in the basis of stationary states.

| Harmonic order, $N$ | $\Omega$ (a.u.) | | $I_0$ (W/cm$^2$) | | Harmonic frequency, $N\Omega$ (a.u.) | |
|---|---|---|---|---|---|---|
| | Expansion over stationary states | 2D model | Expansion over stationary states | 2D model | Expansion over stationary states | 2D model |
| 3 | 0.2659 | 0.2750 | $2.56\times10^{14}$ | $5\times10^{14}$ | 0.7977 | 0.825 |
| 5 | 0.1977 | 0.2354 | $5.179\times10^{14}$ | $10^{15}$ | 0.9885 | 1.177 |
| 7 | 0.1613 | 0.1957 | $7.906\times10^{14}$ | $10^{15}$ | 1.129 | 1.370 |
| 9 | 0.1419 | 0.1749 | $1.207\times10^{15}$ | $10^{15}$ | 1.277 | 1.574 |

At the same time, there are also significant differences between the results of the models used. In particular, Fig. 3 reveals a resonant enhancement of the harmonic yield in the $E_N^{(exp)}(\Omega,I_0)$ dependences in the vicinity of the unperturbed boundaries of multiphoton transitions, $\Omega=I_p/M$ (cf. (17)), which is absent in the $E_N^{(2D)}(\Omega,I_0)$ dependences (Fig. 4). Such a resonant enhancement in the harmonic energies $E_N^{(exp)}(\Omega,I_0)$, calculated via wave-function expansion over the stationary states, is explained by the population of the Rydberg states of the atom, the excitation amplitudes of which (due to the high polarizability of these states) contain a significant number of harmonics of the laser field, which leads to the appearance of additional harmonic generation channels in the spectrum of the dipole acceleration of the atom (12) (an increase in the number of substantially nonzero terms in the sum (14)). As mentioned above, in the calculations based on the 2D one-electron model, because of the limited spatial grid, the highly excited atomic states are distorted (their size and dipole moments of transitions between them are underestimated), which can explain the absence of such regions of resonant enhancement in the energies of harmonics $E_N^{(2D)}(\Omega,I_0)$, calculated in the 2D one-electron model of the helium atom.

On the other hand, the occurrence of resonance peaks in the $E_N^{(exp)}(\Omega,I_0)$ dependences in the vicinity of unperturbed boundaries of multiphoton transitions, $\Omega=I_p/M$, indicates an underestimation of the Stark effect in the model based on the wave-function expansion over the basis of stationary states (8)-(10). In the 2D one-electron model, the ionization potential and the energies of atomic excited states in the laser field shift relative to the unperturbed values by an amount of the order of the electron ponderomotive energy (see Eqs. (17) and (18), and also Figs. 2 and 4). At the same time, in the model (8)-(10), the shift of the state energies due to the Stark effect is qualitatively correctly reproduced only for the ground and lower excited states, while the shift of the energies of highly excited states and the atomic ionization potential by the value of the electron ponderomotive energy does not occur. This is due to the use of a limited basis of stationary states and neglect of the recombination of the active electron [34]. As a result, in model (8)-(10), the boundaries of multiphoton transitions of different orders have an unperturbed position $\Omega_p^{(M)} = I_p/M$ (black vertical lines in Fig. 3).

The noted features are also observed in the dependence of the final probability of atomic ionization on the frequency and peak intensity of the laser field, $W_{ion}^{(exp)}(\Omega,I_0)$, shown in Fig. 5, which were calculated by wave-function expansion over stationary states. The resonant enhancement of the probability of atomic ionization is observed at laser field frequencies $\Omega$ that are lower than or of the order of the unperturbed boundary of $M$-photon transitions, $I_p/M$ (black vertical lines in Fig. 5). In addition, from a comparison of Fig. 5 with Fig. 2, which shows a similar dependence $W_{ion}^{(2D)}(\Omega,I_0)$ calculated in the 2D one-electron model of the helium atom, two more conclusions can be made. First, in the high-frequency region, $\Omega>I_p/2$, the 2D one-electron model overestimates the ionization probability due to the presence in the effective one-electron potential (5) of quasi-stationary states located directly above the ionization potential, which are absent in the real 3D two-electron helium atom. Second, the 2D one-electron model underestimates the role of multiphoton resonances in a relatively low-frequency laser field. This circumstance is apparently also explained by the distortion (underestimation of spatial dimensions) of highly excited states, as well as by the fact that with a decrease in the field frequency (and an increase in the order of resonances), the probability of populating precisely highly excited states of a real atom increases.

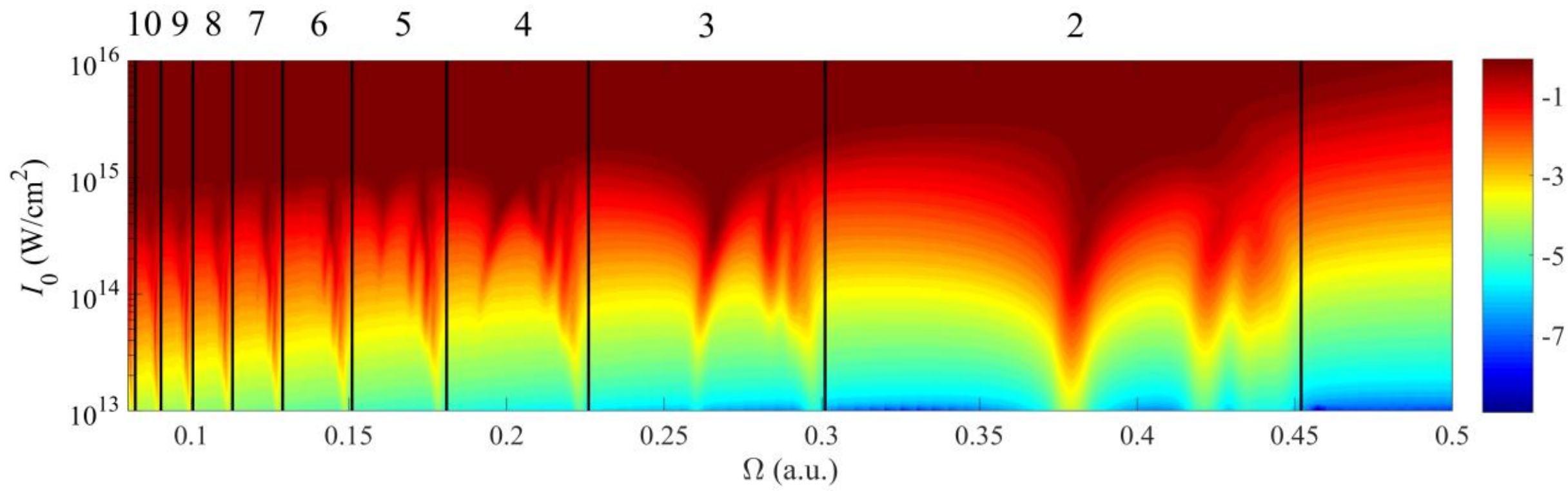

FIG. 5. Dependences of the decimal logarithm of the probability of ionization of a helium atom, $\log_{10}(W_{ion}^{(\exp)})$, at the end of the laser pulse (1), (2) on its frequency Ω and peak intensity $I_0$, obtained by expanding the atomic wave function in the basis of stationary states. Black lines indicate the estimates of the boundaries of multiphoton transitions of orders $M = 2,\ldots,11$, determined with the unperturbed value of the ionization potential of the He atom, $\Omega_p^{(M)} = I_p/M$. The numbers above indicate the regions of $M$-photon transitions.

Thus, the models used are in qualitative agreement with each other in the ranges of Ω and $I_0$ values corresponding to the optimal conditions for generating harmonics of a short-wave laser field of moderate order ($N$ = 3,5,7,9). The model based on the wave-function expansion over stationary states turns out to underestimate the Stark effect, as well as the optimal values of the laser field frequency and the frequencies of the generated harmonics due to neglect of the recombination of the active electron and taking into account the limited number of stationary states. In turn, the differences in the optimal values of the field peak intensity are due to the limitations of both models, in particular, the distortion of the characteristics of highly excited states in the 2D one-electron model, as well as differences in the description of the atomic ionization process and the different intensity steps.

## V. THE INFLUENCE OF THE ORDER OF RESONANT TRANSITIONS ON THE EFFICIENCY OF HARMONIC GENERATION

Let us return to the analysis of the dependences of the harmonic energies on the frequency and peak intensity of the laser field, shown in Figs. 1, 3, and 4. As follows from these figures, with an increase in the peak intensity of the field, the optimal frequency of the field also increases. This tendency is observed for harmonics of all considered orders (from the 3rd to the 9th) and each of the calculation methods. The optima in the above-mentioned figures shift from the left below to the right above. This means that with an increase in the field intensity, the optimal order of resonance, at which the given harmonic is generated, decreases. To analyze this tendency, let us turn to Figs. 3 and 4. For each of the models, we will use those boundaries of multiphoton transitions of different orders that are inherent in this particular model (black vertical lines in Fig. 3 and black curves in Fig. 4).

As an example, let us consider the generation of the third harmonic (Figs. 3(a) and 4(a)). The absolute maximum in the energy of the third harmonic in both models is reached in the condition of the three-photon resonance with the transition $|1s^2\rangle \leftrightarrow |1s2p\rangle$ [30]. However, for laser field intensities exceeding the optimal value, the maximum in the energy of the third harmonic shifts to the region of two-photon resonances. Similarly, the absolute maximum in the energy of the fifth harmonic (Figs. 3(b) and 4(b)) is reached in the region of four-photon resonances. However, for high intensities of the laser field, the maximum energy of the fifth harmonic shifts to the region of three-photon and then (for the highest values of intensity considered in model (8)-(10)) two-photon resonances. The same tendency is observed for the 7th and 9th harmonics.

It is noteworthy that in both models, the optimum for generating the 3rd harmonic corresponds to a three-photon resonance, but the optimum for generating the 5th harmonic is reached under the condition of a four-photon rather than a five-photon resonance. The optimum for the 7th harmonic corresponds to a five-photon resonance, and the optimum for the 9th harmonic corresponds to a six-photon resonance. Thus, the higher the harmonic order *N*, the more the optimum multiphoton resonance order *M* differs from the value of *N* at the optimal laser intensity. At the same time, at low intensity values for all the harmonics considered, the optimum value of *M* approaches and often reaches *N*.

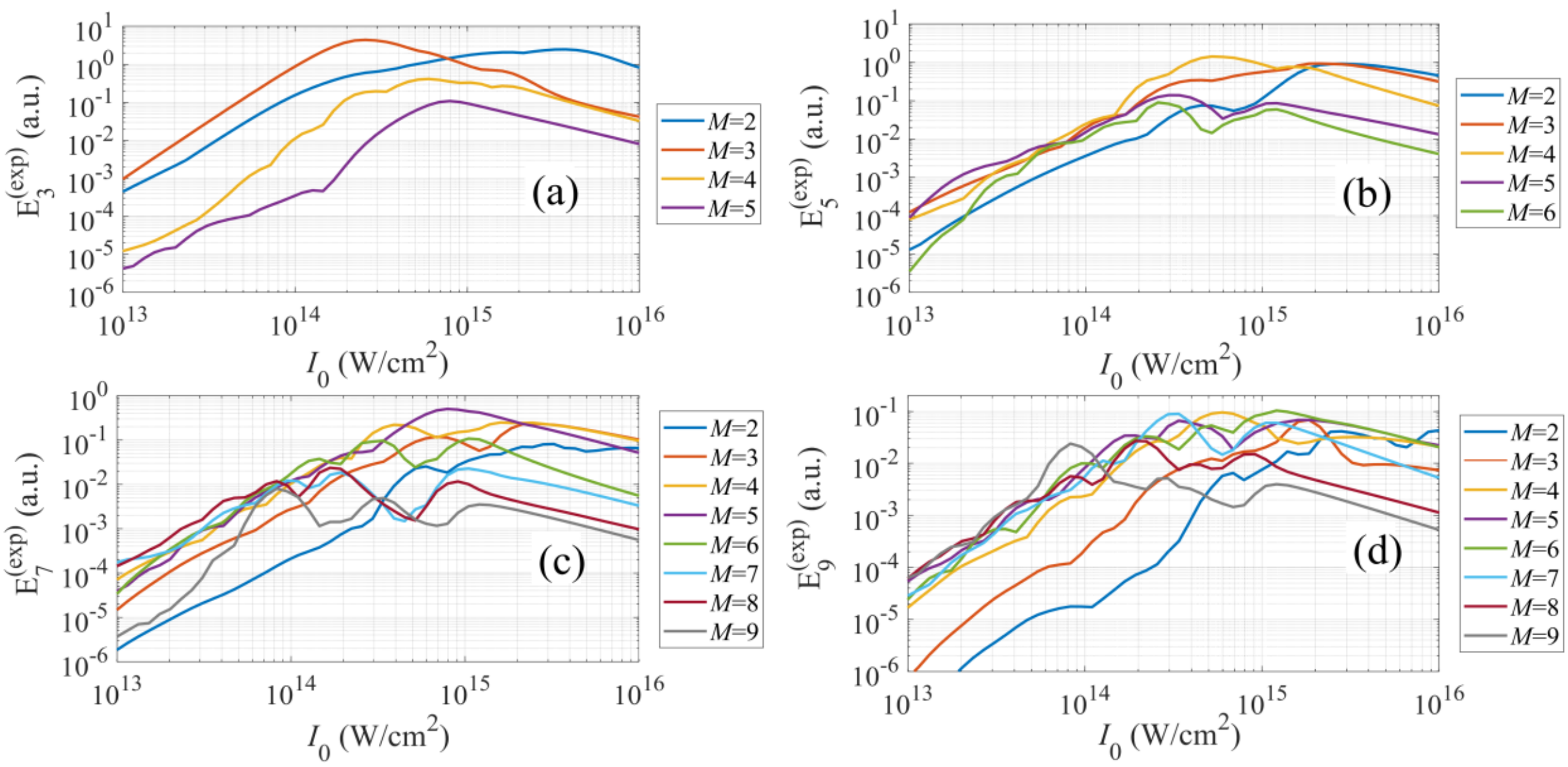

FIG. 6. Dependences of the maximum energies of (a) 3rd, (b) 5th, (c) 7th, and (d) 9th harmonics on the laser peak intensity $I_0$ for the field frequencies in the regions of multiphoton transitions of different orders *M*, for the model (8)-(10) based on the expansion of the atomic wave function over the basis of stationary states. Blue, red, mustard, purple, green, cyan, burgundy, and gray colors correspond to the resonance orders *M* = 2, 3, 4, 5, 6, 7, 8, and 9, respectively.

In terms of the influence of the multiphoton resonance order on the energy of moderate-order harmonics, the results of both models are close and often the same. Let us analyze the results obtained in the model with the wave-function expansion over the basis of stationary states of the helium atom, given in Fig. 6, which shows the dependences of the maximum achievable energy of (a) 3rd, (b) 5th, (c) 7th, and (d) 9th harmonics of the laser field on its intensity for the field frequency in the region of multiphoton transitions of different orders $2 \leq M \leq 9$.

Several conclusions can be drawn from Fig. 6. First, at a low intensity of the laser field, the optimal way to generate the $N$th harmonic is to use multiphoton resonances of order $M$ comparable with the harmonic order $N$, but not exceeding it, $M \leq N$. If the order of resonance is considerably lower, $M \ll N$, or, conversely, higher than the harmonic order, $M > N$, the harmonic energy decreases significantly. This behavior can be interpreted as follows. At a fixed laser intensity, the probability of multiphoton excitation of an atom increases with a decrease in the order of resonance $M$, which makes harmonic generation under conditions of multiphoton resonances, the order of which exceeds the harmonic order, $M > N$, energetically unfavorable. On the other hand, for $M < N$, in order to generate the $N$th harmonic, in addition to the resonant $M$-photon excitation of the atom, non-resonant interaction with $(N - M)$ photons of the laser field is required, the probability of which decreases with increasing value of $(N - M)$. As a result, for a low field intensity and for small values of the difference $(N - M)$, the increase in the probability of resonant excitation of the atom with decreasing $M$ is compensated by a decrease in the probability of non-resonant $(N - M)$-photon interaction. Accordingly, resonances of orders $M = N$ and $N/2 < M < N$, in a low-intensity field give comparable energy of the $N$th harmonic. Note that if the goal is not to maximize the energy of the $N$th harmonic in the spectrum of the dipole acceleration of the atom, but to generate it efficiently at low atomic ionization level, the use of resonances of different orders leaves variability in the choice of the laser field frequency. Note also that for $M \geq 6$, the boundaries of resonances of different orders cannot be determined unambiguously, since the frequencies of $M$-photon transitions to lower excited atomic states and $(M + 1)$-photon transitions to Rydberg states overlap.

Secondly, for resonances of arbitrary order $M$, the maximum energy of a harmonic of arbitrary order $N$ first increases with increasing laser field intensity, then reaches its maximum value and subsequently decreases. The growth of the harmonic energy at low field intensity is due to the fact that in this case ionization practically does not prevent multiphoton excitation of the atom. As a consequence, the efficiency of harmonic generation, determined by the product of the corresponding Fourier components of the amplitudes of the ground and excited states of the atom (see Eq. (14)), increases. With increasing laser field intensity, the excited states of the atom are increasingly depleted due to growing ionization. As a result, at a certain field intensity, the maximum harmonic energy for the considered resonance order is achieved, and with a further increase in intensity, the harmonic energy decreases, as it is shown for the 3rd harmonic in [30]. Note that in the general case, the described dependence of the harmonic energy on the laser field intensity is superimposed with quasiperiodic oscillations caused by a change of the resonant state of the atom, the excitation of which leads to the most efficient harmonic generation at a given laser field intensity and resonance order.

Thirdly, the smaller the order of resonance $M$, the higher, with some exceptions, the corresponding optimal value of the laser intensity for generating a harmonic of order $N$, and vice versa, the higher the intensity, the lower the optimal value of the order of resonance. This behavior is explained by the growth of the efficiency of multiphoton excitation of the atom with a decrease in the order of resonance $M$ and, as a consequence, an increase in the field intensity necessary for the depletion of excited states.

Fourthly, the higher the harmonic order $N$, the more the optimal order of resonance $M$ differs from $N$ at the global (in intensity and frequency of the laser field) maximum of its energy. This regularity can be explained as follows. The higher the harmonic order $N$, the lower the frequency of the laser field must be so that the photon energy of the $N$th harmonic is comparable with the excitation energies of the bound states, and resonant enhancement can take place. In addition, with an increase of the harmonic order $N$, the optimal intensity of the laser field also increases so that the atomic response becomes more nonlinear. The increase in intensity and decrease in frequency of the field lead to the fact that at a fixed order of resonance $M$, the probability of non-resonant interaction with ($N$-$M$) photons of the laser field with subsequent emission of the $N$th harmonic increases. In combination with the previously noted increase in the probability of resonant excitation of an atom with a decrease in the order of resonance $M$, this explains the observed effect.

Note also that if the order of resonance coincides with the harmonic order, $M = N$, then the highest harmonic energy is reached under conditions of maximization of quantum coherence at the resonant transition, i.e. at the maximum value of the product of the central Fourier components of the excitation amplitudes of the resonant states (13a). For instance, to achieve the highest energy of the third harmonic, it is necessary to maximize the population of the $|1s2p\rangle$ state without excessively depleting the ground state $|1s^2\rangle$ [30]. At the same time, if the order of resonance is less than, or, conversely, greater than the harmonic order, $M < N$ or $M > N$, it is necessary to maximize the sum of the products of the side Fourier components of the excitation amplitudes of the resonance states (13a). In this case, to maximize the energy of the $N$th harmonic, it is necessary not only to populate the resonant excited state of the atom, but also to transform it into a multi-frequency "field dressed" atomic state, the spectrum of which should be the wider, the greater the absolute value of the difference $|N - M|\Omega$.

## VI. HARMONIC SPECTRA AND TIME-DEPENDENCES. POPULATION DYNAMICS OF STATIONARY STATES UNDER OPTIMAL CONDITIONS

Let us turn to the dipole acceleration spectra generated under conditions of maximization of the energy of the 3rd, 5th, 7th, or 9th harmonic. These spectra are shown in Fig. 7. Several conclusions follow from them. Firstly, for optimal parameters of the laser field (see Table 1), the spectra of moderate-order harmonics in the induced dipole acceleration, calculated using both models via Eqs. (6) and (12), are qualitatively similar. This means that under optimal conditions, the main mechanism for generating moderate-order harmonics of the short-wave field are bound-bound transitions, since other mechanisms in model (8)-(10) with the wave-function expansion over the stationary states are not taken into account. This is in contrast to the preceding works on the generation below-threshold and near-threshold harmonics by the IR field [14, 15, 17, 19, 20, 21, 23, 27], where rescattering of a released electron (the Corkum's mechanism) plays an important role. This difference is due to the different wavelength of the laser field of the fundamental frequency and the corresponding difference in the multiplicities of multiphoton transitions. For a long-wave IR field, atomic excitation loses frequency selectivity and the role of multiphoton resonances decreases. However, as we have shown, the maximum energy of the 3rd, 5th, 7th and 9th harmonics is achieved for the generating field of the UV and even VUV range. For such a short-wave field, the photon multiplicity of resonances decreases several times vs. the IR field, which greatly increases the efficiency of excitation of bound states and makes bound-bound transitions the dominant mechanism for the moderate-order harmonic generation.

Secondly, in both models for the laser field parameters maximizing the energy of the $N$th harmonic, $3 \le N \le 9$, this harmonic dominates in the dipole acceleration spectrum. At the same time, the amplitudes of the remaining harmonics increase as the harmonic order approaches $N$. In particular, under conditions maximizing the yield of the 9th harmonic (Fig. 7d), the harmonic amplitudes increase upon transition from the 3rd to the 5th harmonic and further to the 7th and 9th ones, which indicates a substantially nonperturbative regime of their generation. At the same time, the squares of the modulus of amplitude of the 3rd, 5th, 7th, 9th, and 11th harmonics differ by slightly more than an order of magnitude, which makes it possible to form from them a train of attosecond pulses with a duration much shorter than the driving field period, see Figs. 8 and 9.

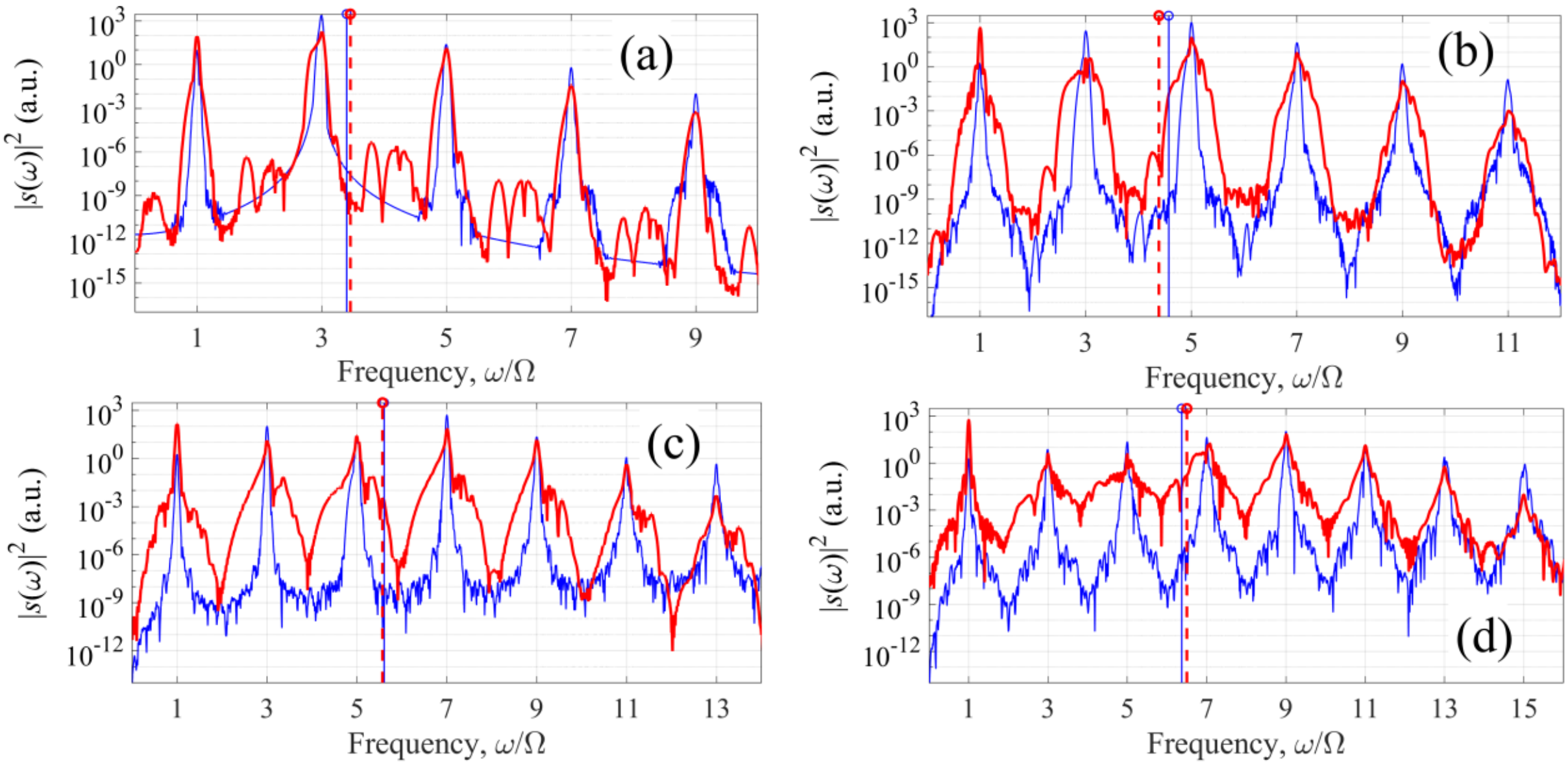

FIG. 7. Dipole acceleration spectra of the helium atom, $|s(\omega)|^2$, calculated in the 2D one-electron model (thick red curve) and by expanding the wave function in stationary states (thin blue curve), under conditions of maximizing the energy of (a) 3rd, (b) 5th, (c) 7th, and (d) 9th harmonics (see Table 1). Blue solid vertical lines correspond to the unperturbed ionization potential of the atom in model (8)-(10), $I_p$ = 0.9036 a.u., normalized to the laser field frequency. Red dashed vertical lines correspond to the normalized ionization potential in the 2D one-electron model of the He atom, $I_p$ = 0.9034 a.u., taking into account its shift by the value of the free-electron ponderomotive energy in the laser field (17).

Thirdly, under conditions maximizing the yield of the 3rd harmonic (Fig. 7a), its amplitude in the dipole acceleration spectrum significantly exceeds the amplitudes of other harmonics and exceeds the amplitude of the atomic response at the frequency of the generating field. As a consequence, the induced dipole acceleration is a quasi-monochromatic VUV signal at the frequency of the 3rd harmonic (see [30]).

Fourthly, Fig. 7 shows the values of the unperturbed ionization potential of the helium atom normalized to the laser field frequency in the model with the wave-function expansion over the basis of stationary states and the ionization potential shifted by the electron ponderomotive energy in the laser field in the 2D one-electron model. As can be seen from Fig. 7, for the laser field parameters that maximize the energy of moderate-order harmonics, these values practically coincide, which confirms the presence of a shift in the frequencies of multiphoton transitions by the electron ponderomotive energy in the 2D one-electron model and the absence of this shift in model (8)-(10).

We also note that the response of an atom at the frequency of the driving field in the 2D one-electron model is significantly greater than in the model with the wave-function expansion in stationary states, which is apparently due to the fact that free-free transitions are not taken into account in the latter model.

Let us now turn to the time dependences of induced dipole acceleration. Figure 8 shows the time dependences of the dipole acceleration, calculated in a 2D one-electron model of the helium atom under conditions that maximize the energy of the 3rd (Figs. 8a and 8b) and 9th (Figs. 8c and 8d) harmonics, after suppression of the fundamental frequency spectral component. Figures 8a and 8b correspond to the resonant three-photon excitation of the state |1s2p⟩, whereas Figs. 8c and 8d correspond to the six-photon excitation of even atomic states (see Sec. V). The respective harmonic spectra are shown in Figs. 7a and 7d. While Figs. 8a and 8c present the time dependences of dipole acceleration on the scale of the laser pulse envelope, Figs. 8b and 8d show the time dependences on the scale of the laser field period.

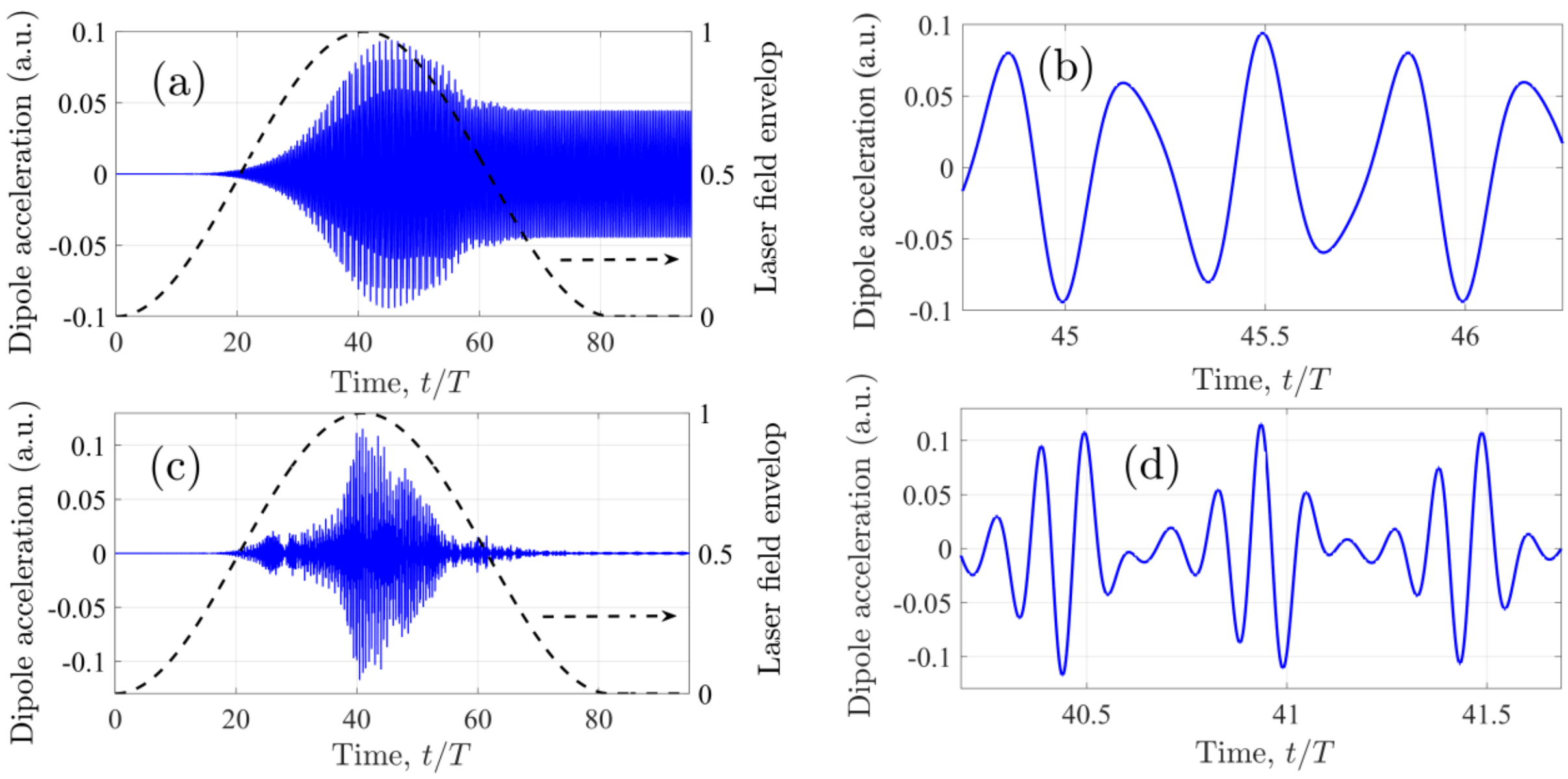

FIG. 8. Dipole acceleration of the helium atom, $\ddot{d}_{2D}(t)$, calculated in the 2D one-electron model (6) (blue solid curve, left vertical axis) after suppression of its fundamental frequency component under conditions of maximizing the energy of 3rd (a), (b), and 9th (c), (d) harmonics (see Table 1). Time is normalized to the laser field period $T = 2\pi/\Omega$. Black dashed curve and right vertical axis in (a) and (c) correspond to the laser field envelope $f(t)$. Panels (b) and (d) show sub-laser-field-cycle dynamics of the dipole acceleration near the peak of its envelope.

As follows from Figs. 8a and 8c, the maximum amplitude of the harmonic signal is achieved in the vicinity of the maximum of the laser field envelope. However, in Fig. 8a, under conditions of the three-photon excitation of the state |1s2p⟩, the time dependence of the dipole acceleration, in addition to the harmonic signal generated in the presence of the laser field, also contains a signal of the extreme ultraviolet free induction decay (XFID) [14, 16, 20, 22, 35] generated at the unperturbed frequencies of dipole-allowed transitions to the ground atomic state (in the case of Fig. 8(a) – at the transition frequency $|1s^2\rangle \leftrightarrow |1s2p\rangle$) after the laser pulse has ended. In Fig. 8c there is no such signal, since under the conditions of six-photon excitation, electric dipole transitions from the resonantly populated states to the ground state of the atom are prohibited.

It worth noting that multiplying the dipole acceleration by the time mask $f(t)$, which is nonzero only in the presence of the laser field, excludes the radiation of XFID from consideration when calculating the energies of harmonics (16). The reason for this is the difference in the spectral, temporal, and spatial properties of XFID radiation and harmonic radiation, in particular, (a) the much longer duration of XFID radiation, determined by relaxation times in the interaction medium, (b) its greater spatial divergence and, in general, (c) the non-multiplicity of the central frequency of XFID radiation to the frequency of the laser field [14, 16, 20, 22, 35]. Meanwhile, the optimal conditions of harmonic generation found in the present paper can be implemented by maximizing the yield of the 5th, 7th, and 9th harmonics with photon energies above the ionization potential of the atom (according to Table 1 and Figs. 7(b)-(d), under the optimal conditions the photon energies of the 5th, 7th, and 9th harmonics exceed the ionization potential). In turn, the third harmonic radiation can be discriminated from the XFID by the different carrier frequency (under optimal conditions, the third harmonic frequency differs from the unperturbed transition frequency), significantly larger bandwidth, and smaller spatial divergence.

Let us now turn to the sub-laser-cycle time dynamics of the dipole acceleration. In agreement with Fig. 7a, the third harmonic dominates in the time dependence of the dipole acceleration in Fig. 8b, while the other harmonics lead to modulation of the quasi-monochromatic signal at its frequency. At the same time, the wide spectrum of near-threshold harmonics generated under conditions of maximizing the 9th harmonic yield, see Fig. 7d, leads to the formation of short beats in the time dependence of the induced dipole acceleration in Fig. 8d after suppression of its fundamental frequency spectral component, which correspond to an attosecond pulse train in the time dependence of the square of the dipole acceleration modulus, shown in Fig. 9.

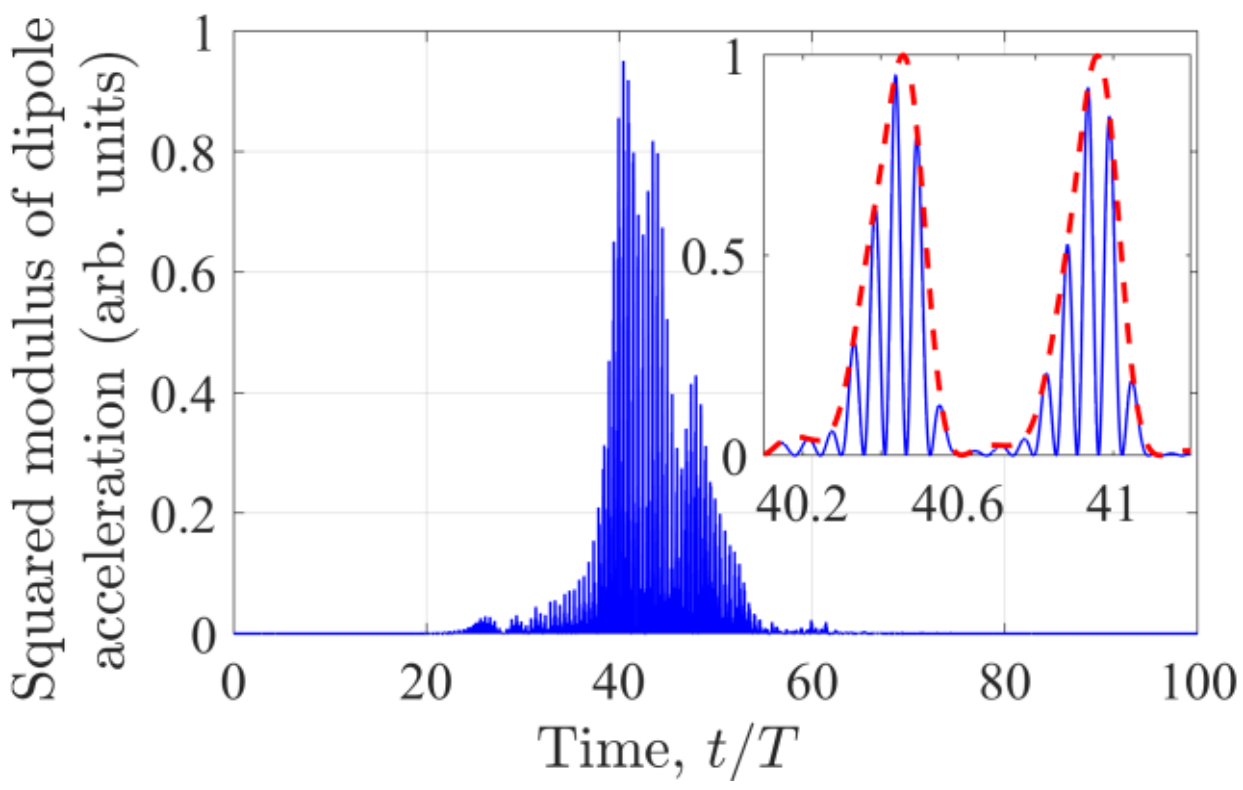

FIG. 9. Squared absolute value of dipole acceleration of the helium atom, $\left|\ddot{d}_{2D}(t)\right|^2$, calculated in the 2D one-electron model (6) (blue solid curve) after suppression of its fundamental frequency component under conditions of maximizing the energy of 9th harmonic (see Table 1, Figs. 7a, 8c, and 8d). Time is normalized to the laser field period $T = 2\pi/\Omega$. Inset shows attosecond bursts in the time dependence of squared modulus of dipole acceleration near the peak of its envelope. Red dashed curve in the inset is the squared absolute value of Hilbert transform of $\ddot{d}_{2D}(t)$. The duration of each burst at FWHM is 130 as and the burst repetition period is 430 as.

As follows from Fig. 9, under conditions that maximize the energy of the 9th harmonic, in the vicinity of the maximum of the envelope of square modulus of dipole acceleration, it represents a train of pulses with a duration of about 130 as and a repetition period of 430 as. Note also that the conditions of Fig. 9 (laser field frequency is 0.1749 a.u. and peak intensity $10^{15}$ W/cm$^2$) can be implemented using the third harmonic of a Ti:Sa laser as the generating field.

Finally, we will analyze the temporal dynamics of the populations of bound states, $|a_k(t)|^2$, and continuum states, $\sum_{l=0}^{L_{max}} \int_0^\infty |b_l(\varepsilon,t)|^2 d\varepsilon$, under conditions that maximize the energy of moderate-order harmonics. This possibility is provided by model (8)-(10), based on the expansion of the wave function of the helium atom over the basis of stationary states. Figure 10 corresponds to the field parameters that maximize the energy of the third harmonic in the spectrum of the atomic dipole acceleration, see Figs. 3a, 7a, 8a and 8b. As follows from this figure, under optimal conditions for generating the third harmonic, at the end of the laser pulse the atom finds itself with a significant probability, 28%, in a resonant excited state $|1s2p\rangle$, with a probability of 31% it is ionized, and with a probability of less than 1/2 (namely, 41%) it remains in the ground state $|1s^2\rangle$. Besides, in the presence of a laser field, other states are noticeably populated, primarily, the $|1s2s\rangle$ state which is the most probable intermediate state in the resonant three-photon transition from $|1s^2\rangle$ state to $|1s2p\rangle$ state. The peak population of the $|1s2s\rangle$ state is 12%.

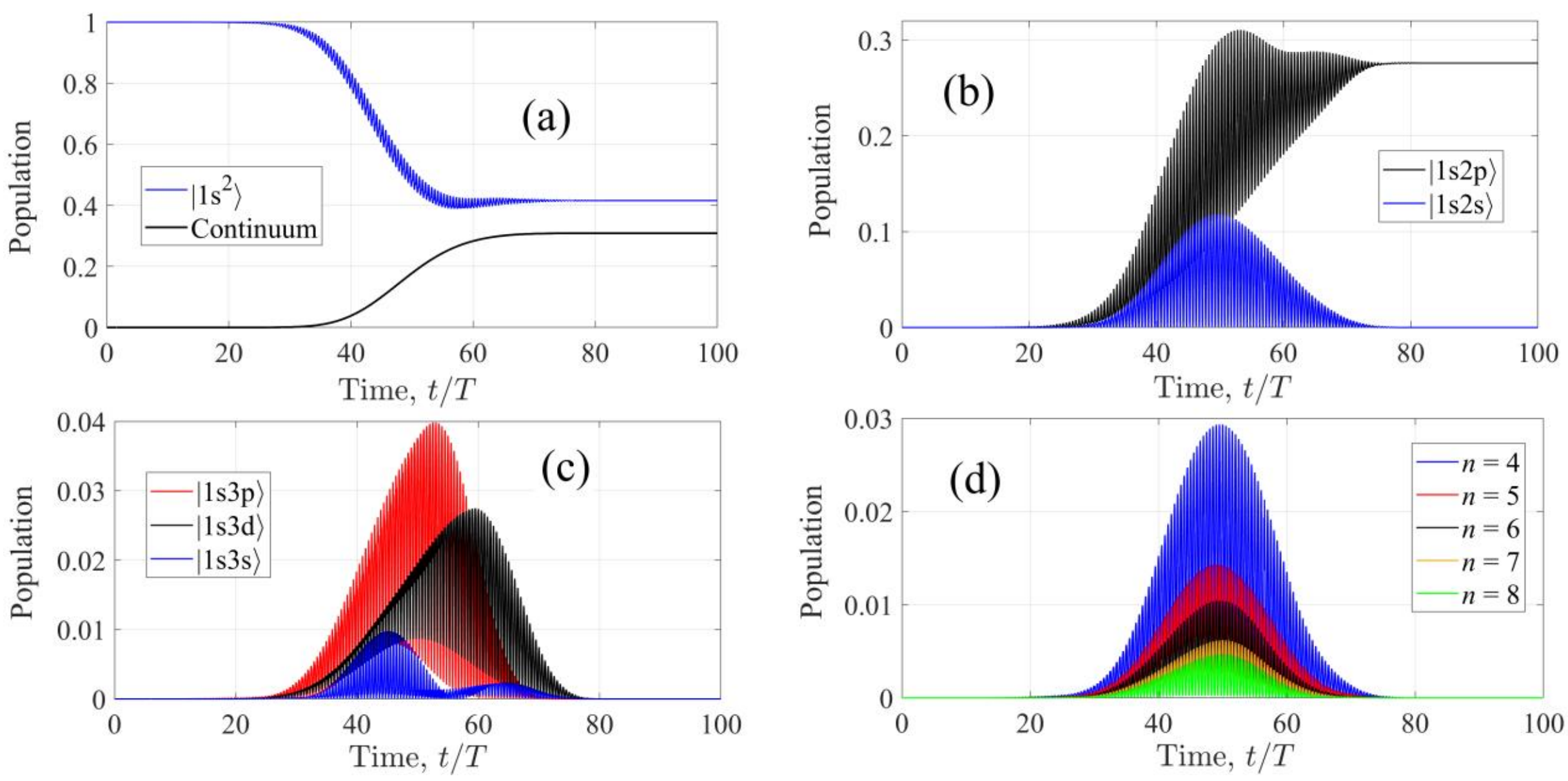

FIG. 10. Time dependences of the populations of the stationary states of the He atom, calculated on the basis of model (8)-(10) taking into account 36 bound states with the principal quantum numbers $n = 1,\ldots,8$. Time is normalized to the laser field period $T$=2π/Ω. Panel (a) shows population of the ground state $|1s^2\rangle$ and total population of the continuum states; panels (b) and (c) – populations of the states with principal quantum numbers $n$=2 and $n$=3. Panel (d) shows sums of the populations of states with the same values of the principal quantum number $4 \le n \le 8$. The figure is drawn for $\Omega = 0.2659$ a.u. and $I_0 = 2.5\times10^{14}$ W/cm$^2$, which corresponds to the optimal conditions for the generation of the third harmonic.

Under conditions that maximize the yield of higher-order harmonics, the final population of atomic excited states turns out to be much smaller, while the ionization probability increases and the population of the ground state decreases, see Fig. 11. This happens for two reasons. Firstly, with an increase in the harmonic order $N$, the optimal order of resonance $M$ also increases (cf. Fig. 6), as a result of which the efficiency of multiphoton excitation of the atom decreases. Secondly, an increase in the intensity and a decrease in the frequency of the laser field lead to an increase in the rate of single-photon ionization (from those stationary states whose ionization potential is smaller than the photon energy of the laser field), which is caused by an increase in the photoionization cross section with a decrease in the electron energy in the continuum, see, for example, Appendix A in [30].

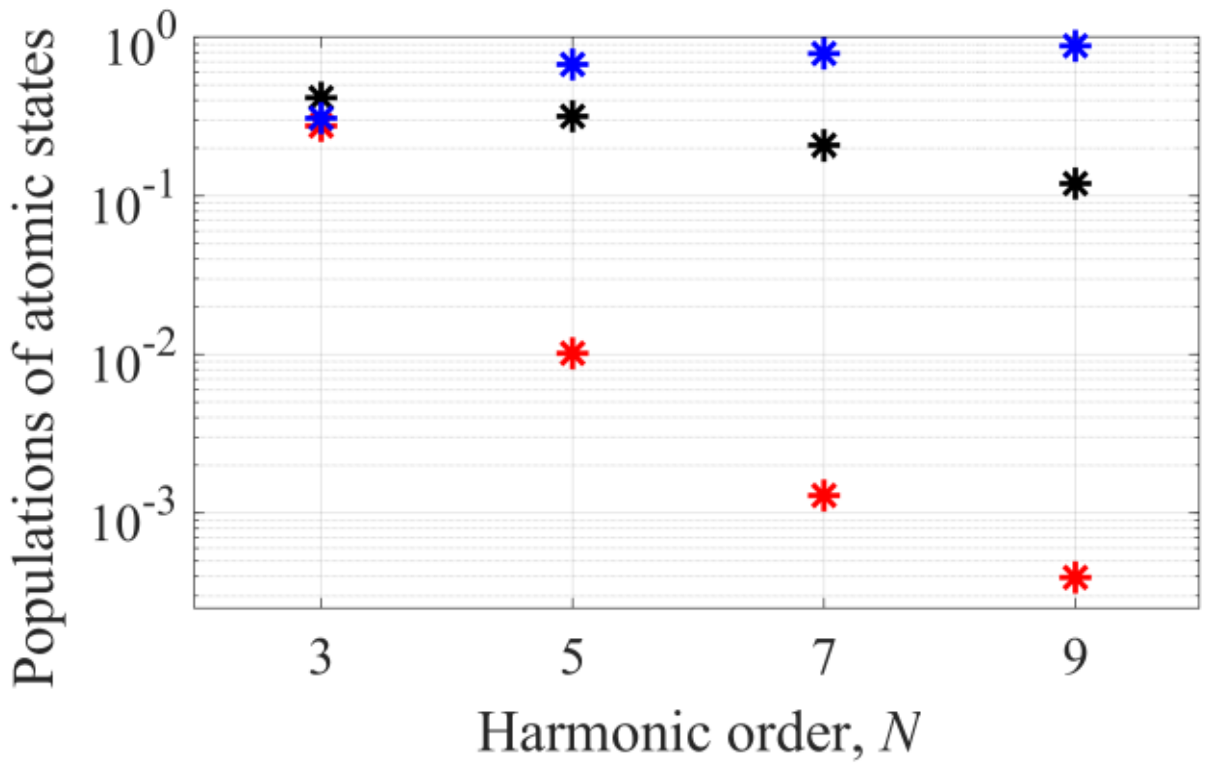

FIG. 11. The population of the ground state (black stars), the total population of the excited states (red stars), and the population of the continuum (blue stars) after the end of the laser pulse depending on the harmonic number $N$ under conditions optimal for its generation (see Table 1).

The above statements are confirmed by Fig. 12, which shows the dynamics of the populations of the atomic stationary states under conditions that maximize the energy of the 9th harmonic in the spectrum of the induced dipole acceleration. In this case, the total final population of the excited states is 0.04%, the final probability of ionization of the atom is approximately 88%, and the final population of the ground state is 12%. Compared to the conditions maximizing the 3rd harmonic energy (Fig. 10), the peak population of excited states in the presence of a laser field also decreases, but the population is distributed over the states more uniformly. It is noteworthy that in this case the time dependences of highly excited states contain several maxima, which may indicate Rabi oscillations between them.

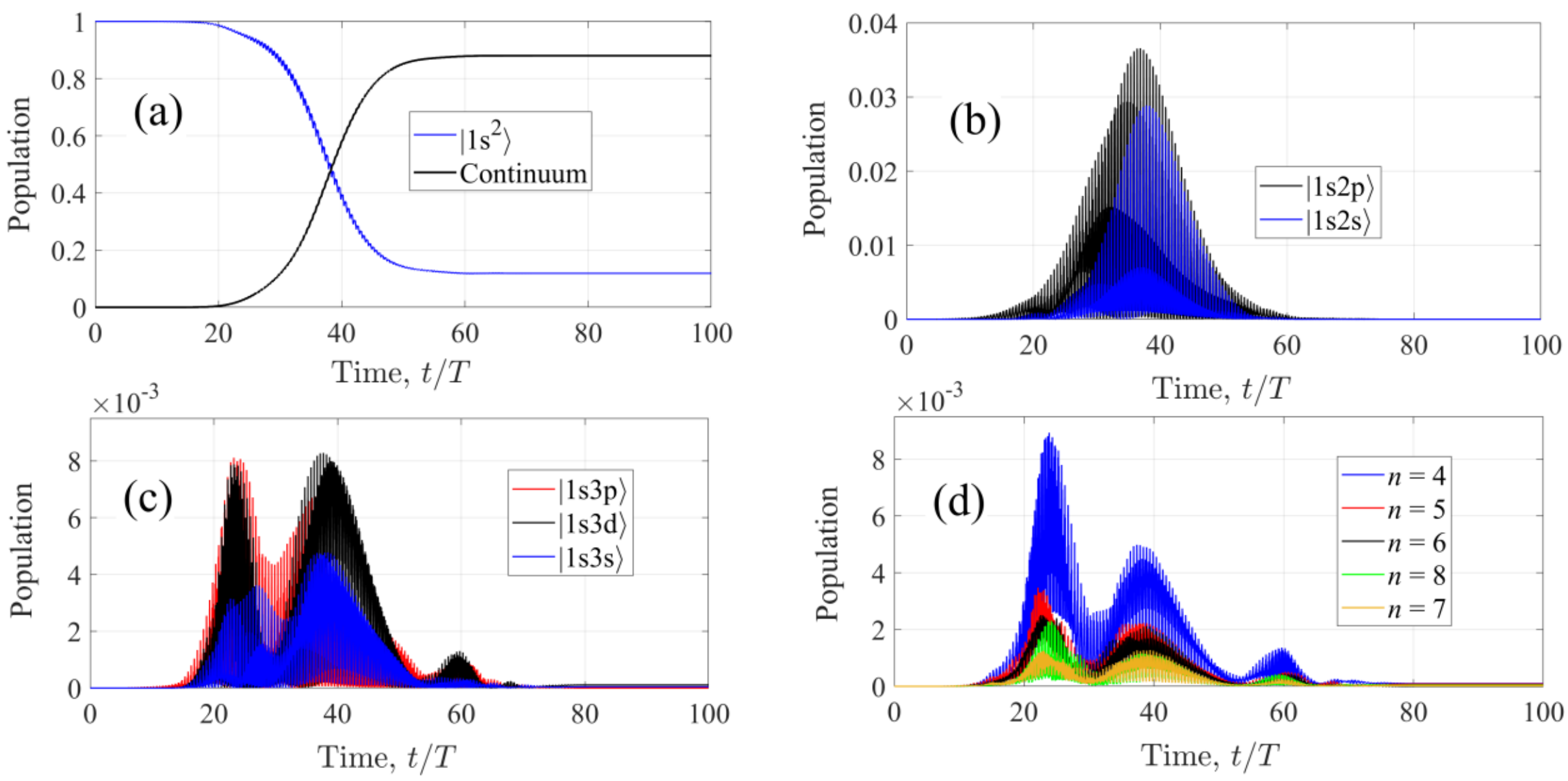

FIG. 12. Same as in Fig. 10, but for optimal conditions for generating the ninth harmonic, $\Omega$ = 0.1419 a.u. and $I_0 = 1.207\times10^{15}$ W/cm$^2$.

## VII. CONCLUSION

In summary, this article analyzes the optimal conditions for generating moderate-order harmonics of a short-wave laser field by helium atoms in a given external field. It is shown that under conditions, which maximize the energy of the 3rd, 5th, 7th, or 9th harmonics of the field, the generation of these harmonics is caused predominantly by the transitions between the bound atomic states, and the maximum energy of the harmonics is reached under the conditions of their resonant multiphoton excitation. It is shown that, depending on the order of the harmonic whose energy is maximized, the optimal frequency of the laser field lies in the UV or even VUV range and is in three-, four-, five-, or six-photon resonance with transitions from the ground to the excited states of the helium atom. In this case, the higher the harmonic order $N$, the more the optimal order of resonance $M$ differs from the value of $N$ at the optimal laser field intensity. In turn, the optimal value of the field peak intensity varies from $2.5\times10^{14}$ to $1.2\times10^{15}$ W/cm$^2$, and the total probability of excitation and ionization of the atom at the end of the laser pulse under optimal conditions exceeds 50%. It is shown that in a sufficiently low-frequency field (such as the second harmonic of a titanium-sapphire laser) at an optimal value of its intensity for generating moderate-order harmonics, the change in the frequencies of multiphoton transitions as a result of the quadratic Stark effect is comparable with their unperturbed values. Accordingly, there is no need for frequency tuning of the field, since changing its intensity allows tuning the field in a multiphoton resonance with an arbitrary excited state of the atom, and even changing the order of resonance for the transition.

It is shown that with increasing intensity of the laser field, the optimal order of resonance decreases. At the same time, in a field of moderate intensity (under conditions of weak ionization of the atom) the use of resonances of orders $M = N$ and $N/2 < M < N$ results in comparable energy of the $N$th harmonic, which leaves variability in the choice of the laser field frequency. A qualitative explanation of these regularities is given based on the competition of processes of multiphoton excitation and ionization of the atom, as well as the balance of probabilities of resonant and non-resonant processes.

It is shown that for the laser field parameters maximizing the energy of the $N$th harmonic, $3 \leq N \leq 9$, this harmonic dominates in the dipole acceleration spectrum, while the amplitudes of the remaining harmonics increase as the harmonic order approaches $N$. In particular, under conditions maximizing the yield of the 9th harmonic, the harmonic amplitudes increase upon transition from the 3rd harmonic to the 5th and then to the 7th and 9th, which indicates a substantially non-perturbative regime of their generation. In this case, the squares of modulus of the amplitude of the 3rd, 5th, 7th, 9 th, and 11th harmonics differ by slightly more than an order of magnitude, which results in formation of a train of pulses with a duration of 130 as and a repetition period of 430 as in the time dependence of squared modulus of dipole acceleration after suppression of its fundamental frequency spectral component.

At the same time, under conditions maximizing the yield of the 3rd harmonic, its amplitude in the dipole acceleration spectrum significantly exceeds the amplitudes of other harmonics and exceeds the amplitude of the atomic response at the frequency of the driving field. As a result, the induced dipole acceleration is a quasi-monochromatic VUV signal at the third harmonic frequency.

Thus, the most efficient generation of moderate-order harmonics by helium atoms occurs if (a) the harmonic frequencies are comparable to or slightly exceed the ionization potential of the atom, taking into account its shift by the electron ponderomotive energy in the field, (b) the laser field frequency is close to the frequencies of the multiphoton transitions to the bound excited states of the atom with an order of resonance equal to or slightly lower than the harmonic order, and (c) the field intensity is sufficient to excite and ionize the majority of the atoms while maintaining a noticeable population in the ground state.

It is worth noting that the use of UV and VUV fields not only significantly increases the single-atom response compared to the case of longer-wavelength drivers, but also ensures better phase matching when generating harmonics in an extended medium at high ionization levels (see, for example, [36]). The study of propagation effects is beyond the scope of this work. For recent reviews on this topic see [37, 38]; the macroscopic aspects of near-threshold harmonic generation by near-IR field are considered in [14, 16, 20, 22, 26]. We can only state that the radiation of a very thin medium will be close in its spectral and temporal properties to the radiation of a single atom. With an increase in the thickness of the medium, the optimal conditions for harmonic generation will probably shift to the region of laser parameters corresponding to a lower degree of atomic ionization, i.e. lower field intensity and/or larger detuning from the frequencies of multiphoton transitions, which will ensure better phase matching of the laser field and its harmonics. However, the difference with the case of a given field can be reduced under conditions of implementation of quasi-phase matching of the generated and generating fields [37].

Finally, due to the similarity of noble gas atoms, the obtained results can be extrapolated to other noble gases with a corresponding decrease in the frequency and intensity of the driving field with a decrease in the ionization potential of the atom.

## ACKNOWLEDGMENTS

This work was supported by the Russian Science Foundation (project No. 24-12-00461).

The authors are grateful to the Joint Supercomputer Center of RAS for the provided supercomputer sources.

---